\documentclass{aa}

\usepackage{graphicx}
\usepackage{wasysym}

\def\comment#1{\noindent{\bf }}

\def\wig#1{\mathrel{\hbox{\hbox to 0pt{%
          \lower.6ex\hbox{$\sim$}\hss}\raise.4ex\hbox{$#1$}}}}

\def\gradad{\nabla_{\rm ad}}
\def\teff{T_{\rm eff}}

\def\fabs{F_{\tiny\RIGHTcircle}}

\def\tday{T_{\rm day}}
\def\tnight{T_{\rm night}}
\def\psday{P^\star_{\rm day}}
\def\psnight{P^\star_{\rm night}}

\def\tsyn{\tau_{\rm syn}}
\def\sec{\rm\, sec}
\def\m{\rm\, m}
\def\kg{\rm\,kg}
\def\ergscm{\rm\,erg\,s^{-1}\,cm^{-2}}
\def\W{\rm\,W}
\def\K{\rm\,K}
\def\km{\rm\,km}
\def\cm{\rm\,cm}
\def\J{\rm\, J}
\def\erg{\rm\, erg}
\def\radius{R}   
\def\orbit{a}    
\def\gasconst{{\cal R}}
\def\oms{\omega_{\rm s}}

\def\hotjups{Pegasi planets}    
\def\hotjup{Pegasi planet}     
\def\hotjov{Pegasi-planet}     
\def\s{$\,\,$}                 
\def\apj{ApJ}

\def\apjl{ApJL}
\def\nature{Nature}

\def\icarus{Icarus}
\def\jas{J. Atmos. Sci.}

\def\science{Science}

\def\Tdn{\Delta T_{\rm day-night}}

\def\Thor{\Delta T_{\rm horiz}}
\begin{document}


\thesaurus{08}
\title{Atmospheric Circulation and Tides of ``51\,Peg\,b-like'' Planets}
\author{Adam P. Showman\inst{1}
\and Tristan Guillot\inst{2}} 
\institute{University of Arizona, 
Department of Planetary Sciences and Lunar and Planetary Laboratory,
Tucson, AZ 85721; showman@lpl.arizona.edu
\and Observatoire de la C\^ote d'Azur, Laboratoire Cassini, CNRS
UMR 6529, 06304 Nice Cedex 4, France; guillot@obs-nice.fr}
\date{Submitted 23 March 2001; Accepted 18 January 2002}
\titlerunning{Dynamics of \hotjov\s Atmospheres}
\authorrunning{A.P. Showman and T. Guillot}
\maketitle

\begin{abstract}
We examine the properties of the atmospheres of extrasolar giant
planets at orbital distances smaller than 0.1\,AU from their stars. 
We show that these ``51\,Peg\,b-like'' planets are rapidly synchronized by tidal
interactions, but that small departures from synchronous 
rotation can occur because of fluid-dynamical torques within these planets. 
Previous radiative-transfer and evolution models of such planets
assume a homogeneous atmosphere.  Nevertheless, we show using simple
arguments that, at the photosphere,
the day-night temperature difference and characteristic wind speeds may 
reach $\sim500\K$ and $\sim2\km\sec^{-1}$, respectively.  Substantial
departures from chemical equilibrium are expected.  The 
cloud coverage depends sensitively on the dynamics; clouds could exist
predominantly either on the dayside or nightside, depending on the
circulation regime.  Radiative-transfer 
models that assume homogeneous conditions are therefore
inadequate in describing the atmospheric properties of 51\,Peg\,b-like planets.
We present preliminary three-dimensional, nonlinear 
simulations of the atmospheric circulation of HD209458b that
indicate plausible patterns for the circulation and
generally agree with our simpler estimates.
Furthermore, we show that kinetic energy production in the atmosphere
can lead to the deposition of substantial energy in
the interior, with crucial consequences for the evolution of these
planets.  Future measurements
of reflected and thermally-emitted radiation from these planets
will help test our ideas.

\end{abstract}

\section{Introduction}

The discovery of extrasolar planets has led to a
growing list of work devoted to modeling their
atmospheres (Burrows et al. \cite{Bur97}; Seager and Sasselov 
\cite{SS98}, \cite{SS00}; Goukenleuque et al. \cite{Gouk00}; 
Barman et al. \cite{Bar01}).
While no spectra of these objects have yet been measured,
one might be encouraged by the successes obtained in the case of
the similar brown dwarfs, for which theoretical models now
reproduce the observations well, even in the case of low-temperature
objects ($\teff\sim 1000\,$K or less) (e.g. Marley et al. \cite{Mar96};
Allard et al. \cite{All97}; Liebert et al. \cite{Lie00}; Geballe et
al. \cite{Geb01}; Schweitzer et al. \cite{Sch01} to cite only a few).
However, an important feature of extrasolar planets is their
proximity to a star: The irradiation that they
endure can make their atmospheres significantly different than
those of isolated brown dwarfs with the same effective
temperature. This has been shown to profoundly alter the
atmospheric vertical temperature profile (Seager and Sasselov
\cite{SS98}; Goukenleuque et al. \cite{Gouk00}; Barman et
al. \cite{Bar01}). We will show that most importantly, it also affects
the horizontal temperature distribution and atmospheric chemistry so
that the models calculated thus far may fail to provide an adequate
description of the atmospheres of the most intensely irradiated
planets. Advection has never previously been considered, but it can play a
major role for the composition, temperature,
spectral appearance and evolution of extrasolar planets.

We will focus on the extrasolar planets for which irradiation is the
most important: 51\,Peg\,b-like planets, which we henceforth dub
``\hotjups'' and define as gas giants
orbiting solar-type stars at less than 0.1\,AU. Their importance
is demonstrated by the fact that they orbit
nearly 1\% of stars surveyed so far and constitute 
27\% of currently-known extrasolar giant planets.  They
are also more easily characterized by the transit method than are other
planets, as
indicated by the discovery of the transiting gas giant HD209458b
(Charbonneau et al. \cite{Cal00}; Henry et al. \cite{Hal00}). 
In the preceeding paper (Guillot and Showman
\cite{GS01}, hereafter Paper~I), we showed how the atmospheric
boundary condition governs the evolution of \hotjups. We also
advocated that an additional source of energy is needed to explain the
radius HD209458b, and that this would most likely be provided by the
downward transport and subsequent dissipation of kinetic energy
with a flux of $\sim 1$\% of the absorbed stellar energy.
In this paper, we use the temperature profiles obtained in
Paper~I to evaluate the dynamical state of \hotjov\s atmospheres,
and discuss how dynamics may influence the cloud abundance, chemical
composition, and thermal state (all of which will be amenable to
observation in the near future).  We then present preliminary
numerical simulations of the circulation that indicate plausible
circulation patterns and show
how downward propagation of kinetic energy from the atmosphere 
to the interior can occur.

After reviewing the expected interior structure (Section 2), we
begin in Section~3 with the problem of tides: \hotjups\s 
have been predicted to rotate synchronously (Guillot et al. \cite{Gui96}), 
implying that they always present the same face toward the
star. We argue however that dynamical torques may
maintain the interior in a non-synchronous rotation state, which
has important implications for understanding atmospheric processes.
In section~4, we discuss the probable wind speeds, day-night
temperature differences, and flow geometries, including both 
order-of-magnitude arguments and our numerical simulations.
A summary of the results is provided in section~5.

\section{Interior structure}
\label{sec:int}

\begin{figure}[htb]
\begin{center}
\resizebox{\hsize}{!}{\includegraphics[angle=0]{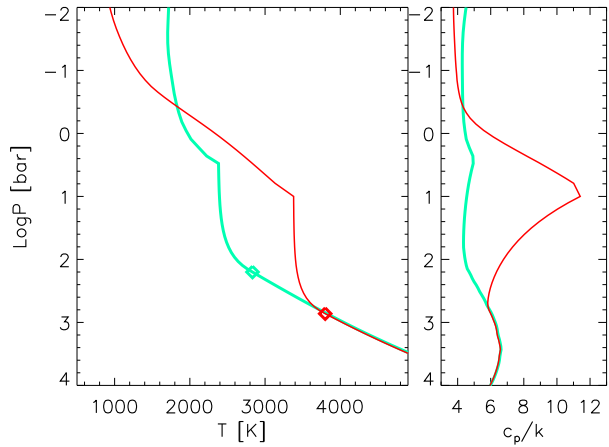}}
\caption{%
{\it Left:} Temperature profiles for the ``hot'' model
(thin line) and ``cold'' model (thick grey line) at
times (5.37 and 0.18 Ga, respectively) when the models
match HD209458b's measured radius.   The diamonds 
indicate the radiative/convective boundary. The terms ``hot'' and ``cold''
are chosen to indicate the models' relative temperatures in the region from 
$\sim1$--300 bars, which is the important region of the atmosphere for
affecting the evolution.  See Paper~I for details. The
discontinuity in the profiles' slopes is unphysical; it results
from the fact that atmospheric models assume a much higher intrinsic
luminosity than predicted by evolution models.  {\it Right:} Dimensionless
mean specific heat per particle in the atmosphere. The local maxima in
$c_p$ are due to the dissociation of the H$_2$ molecule.
}
\label{fig:profiles}
\end{center}
\end{figure}

Fully self-consistent models including both the atmospheres and
interiors of \hotjups\s do not yet exist.  To obtain first-cut
estimates of the expected temperature profiles,
we therefore use the same approach as in
Paper~I: we choose two extreme evolution models (``hot'' and ``cold'')
that match the radius of HD209458b (but at different
ages and hence intrinsic luminosities).  These
evolution calculations only pertain to pressures larger than 3 to 10
bars (depending on the boundary condition), and so we extend the profiles
to lower pressure using radiative-transfer calculations for 
intensely-irradiated planets from Marley 
(personal communication) and Barman (2001). The calculations are not
strictly appropriate to the case of HD209458b, but provide us
with reasonable estimates of the expected structure (and as we will 
see our qualitative
conclusions are not sensitive to the type of profile chosen). The
resulting ``hot'' and ``cold'' models that we will use hereafter are
depicted in Figure~\ref{fig:profiles}. Interestingly, Paper~I shows
that these two models are representative of a relatively wide variety
of models of HD209458b, including those with energy dissipation. 

It should be noted that the high temperatures can
trigger the dissociation of the hydrogen molecule, which is indicated
in Fig.~\ref{fig:profiles} by a local maximum of the heat capacity
$c_p$. This effect is important for our purposes because it implies
that stellar heat can be stored in hot regions and reclaimed in
colder regions by molecular recombination.
Other molecules (e.g. H$_2$O) can undergo dissociation, but this will
be neglected due to their small abundances.

\begin{figure}[ht]
\begin{center}
\resizebox{\hsize}{!}{\includegraphics[angle=0]{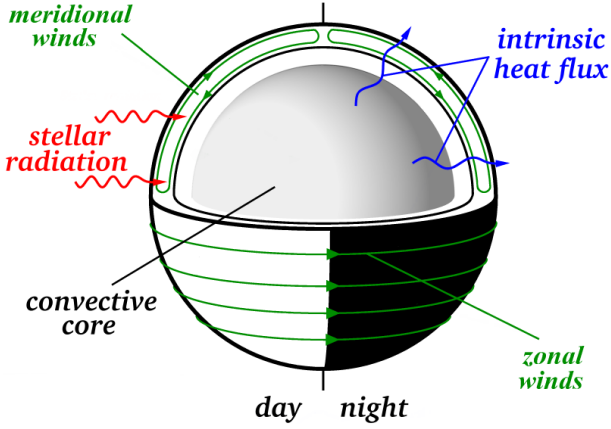}}
\caption{%
Conjectured dynamical structure of \hotjups:
At pressures larger than 100--$800$\,bar, the intrinsic heat flux must
be transported by convection. The convective core is at or near
synchronous rotation with the star and has small latitudinal
and longitudinal temperature variations. At lower pressures a
radiative envelope is present. The top part of the atmosphere
is penetrated by the stellar light on the day side. The spatial
variation in insolation should drive winds that
transport heat from the day side to the night side (see text).}
\label{fig:circ}
\end{center}
\end{figure}

Evolution models of \hotjups\s
show that two interior regions exist: a radiative
zone (including the atmosphere) that extends down to pressures of 
100 to 800\,bar, and a
deeper convective core. The intrinsic luminosity of the planet
is about 10,000 times smaller than the energy absorbed
from the star.  If energy dissipation in the outer layers is 
large (see Paper I), another
convective zone can appear at the levels where dissipation is the
largest. This however requires a very high dissipation of $\sim
10$\% of the incoming heat flux. We therefore chose to only consider
the simple radiative/convective scenario, as depicted in
Fig.~\ref{fig:circ}.

We split the planet into an ``atmosphere'' dominated by stellar
heating, with possible horizontal temperature inhomogeneities, and an
``interior'' including the convective core, for which inhomogeneities
should be much smaller. The crux of the problem is to understand how
the heat absorbed on the day side and near the equator is
redistributed by winds and/or rotation to the night side and to the
poles.

\section{Synchronization of \hotjups}
\label{sec:sync}

It has been shown that the tides raised by the star on \hotjups\s
should rapidly drive them into synchronous rotation (Guillot et
al. \cite{Gui96}; Marcy et al. \cite{Mar97}; Lubow et
al. \cite{Lub97}). This can be seen by considering the
time scale to tidally despin the planet (Goldreich \& Soter
\cite{GS66}; Hubbard \cite{Hub84}):
\begin{equation}
\tau_{\rm syn} \approx Q \left({\radius^3\over GM}\right)(\omega-\omega_{\rm s})
\left({M\over M_\star}\right)^2\left({\orbit\over \radius}\right)^6,
\label{eq:tau-tidal}
\end{equation}
where $Q$, $\radius$, $M$, $\orbit$, $\omega$ and $\omega_{\rm s}$  are
the planet's tidal
dissipation factor, radius, mass, orbital semi-major axis, rotational
angular velocity, and synchronous (or orbital) angular velocity.
$M_*$ is the star's mass,
and $G$ is the gravitational constant. Factors of order unity have
been omitted. A numerical estimate
for HD209458b (with $\omega$ equal to the current Jovian
rotation rate) yields a spindown time $\tau_{\rm syn}\sim 3Q$\,years.
Any reasonable dissipation factor $Q$ (see Marcy et al. \cite{Mar97};
Lubow et al. \cite{Lub97}) shows that HD209458b should be led to
synchronous rotation in less than a few million years, i.e., on a time
scale much shorter than the evolution timescale. Like other \hotjups,
HD209458b is therefore expected to be in synchronous
rotation with its 3.5-day orbital period.

Nevertheless, stellar heating drives the atmosphere
away from synchronous rotation, raising the possibility that the
interior's rotation state is not fully synchronous.
Here, we discuss (1) the energies associated with the planet's
initial transient spindown, and (2) the possible equilibrium
states that could exist at present.

\subsection{Spindown energies}

Angular momentum conservation requires that as the planet spins down, the orbit
expands. The energy dissipated during the spindown process is the
difference between
the loss in spin kinetic energy and the gain in orbital energy:
\begin{equation}
\dot{E}=-{d\over dt}\left({1\over 2}k^2 MR^2\omega^2 
-{1\over 2}M a^2\oms^2 \right),
\end{equation}
where $k$ is the dimensionless radius of gyration ($k^2=I/M\radius^2$,
$I$ being the planet's moment of inertia).  The time derivative is negative,
so $\dot{E}$, the energy dissipated, is positive.
The orbital energy is the sum of the planet's gravitational potential energy and
orbital kinetic energy and is negative by convention.
The conservation
of angular momentum implies that the rate of change of $\oms$ is
constrained by that on $\omega$:
\begin{equation}
{d\over dt}\left(M a^2\oms + k^2 MR^2\omega\right)=0.
\end{equation}
The fact that the planetary radius changes with time may slightly
affect the quantitative results. However, since $\tsyn$ is
so short, it can be safely neglected in this first-order estimate. $R$
being held constant, it is straightforward to show, using Kepler's
third law, that:
\begin{equation}
\dot{E}=-k^2 MR^2 (\omega-\oms)\dot{\omega}.
\end{equation}
(Note that $\dot{\omega}$ is negative, and so $\dot{E}$ is positive.)

The total energy dissipated is $E\approx k^2 MR^2
(\oms-\omega)^2/2$, neglecting variation of the orbital
distance. Using the moment of inertia and rotation rate of
Jupiter ($k^2=0.26$ and $\omega=1.74\times 10^{-4}\,\sec^{-1}$), we
obtain for HD209458b $E\approx 4\times10^{41}\erg$. If
this energy were dissipated evenly throughout the planet, 
it would imply a global temperature increase of 1400\,K.

By definition of the synchronization timescale, the dissipation rate
can be written:
\begin{equation}
\dot E = {k^2 M \radius^2 (\omega - \oms)^2\over \tau_{\rm syn}}.
\end{equation}
With $Q$ of $10^5$, a value commonly
used for Jupiter, $\tsyn\sim
3\times 10^5$ years and the
dissipation rate is then $\sim 10^{29}\erg\sec^{-1}$, or 35,000
times Jupiter's intrinsic luminosity.
Lubow et al. (\cite{Lub97}) have suggested that dissipation in the
radiative zone could exceed this value by up to two orders of magnitude,
but this would last for only $\sim 100$ years.

The thermal pulse associated with the initial spindown is large
enough that, if the energy is dissipated in the planet's interior,
it may affect the planet's radius.  It has previously been argued
(Burrows et al. \cite{Bur00}) that \hotjups\s must have migrated
inward during their first $10^7$\,years of evolution; otherwise, they
would have contracted too much to explain
the observed radius of HD209458b.  But the thermal pulse associated
with spindown was not included in the calculation, and this extra
energy source may extend the time over which migration was possible.
 
Nevertheless, it seems difficult to invoke tidal synchronization
as the missing heat source necessary to explain HD209458b's present (large)
radius. High dissipation rates are possible if $\tsyn$ is small, but
in the absence of a mechanism to prevent synchronization,
$\dot E$ would drop as soon as $t > \tsyn$. The most efficient way of
slowing the planet's contraction is then to invoke
$\tsyn\sim 10^{10}$\,years. In that case, the energy dissipated
becomes $\dot E\sim 10^{24}\erg\sec^{-1}$, which is two orders of
magnitude smaller than that necessary to significantly affect the
planet's evolution (Paper I; Bodenheimer et al. 2001). For the
present-day dissipation to be significant, an initial
rotation rate 10 times that of modern-day Jupiter would be needed.
But the centripetal acceleration due to rotation exceeds the
gravitational acceleration at the planet's surface for rotation
rates only twice that of modern-day Jupiter, so this possibility is
ruled out.  Furthermore, such long spindown times
would require a tidal $Q$ of $\sim 10^9$--$10^{10}$,
which is $\sim 10^4$ times the $Q$ values inferred for Jupiter,
Uranus, and Neptune from constraints on their satellites' orbits
(Peale \cite{Pea99}, Banfield and Murray \cite{Ban92}, Tittemore and 
Wisdom \cite{Tit89}).  
Dissipation of the energy due to transient loss
of the planet's initial spin energy therefore cannot provide the
energy needed to explain the radius of HD209458b.

Another possible source of energy is through circularization of the
orbit.  Bodenheimer et al. (\cite{Bod01}) show that the resulting
energy dissipation could reach $10^{26}\erg\sec^{-1}$ if the planet's
tidal $Q$ is $10^6$ and if a hypothetical companion planet pumps HD209458b's
eccentricity to values near its current observational upper limit of 0.04.
If such a companion is absent, however, the orbital circularization time is
$\sim10^8$ years, so this source of heating would be negligible at present.
Longer circularization times of $10^9$--$10^{10}$ years would allow the heating
to occur until the present-day, but its magnitude is then reduced to
$10^{25}\erg\sec^{-1}$ or lower, which is an order of magnitude smaller
than the dissipation required.

\subsection{The equilibrium state}

The existence of atmospheric winds implies that the atmosphere is
not synchronously rotating.  Because dynamics can transport angular
momentum vertically and horizontally (including the possibility of
downward transport into the interior), the interior may evolve to
an equilibrium rotation state that is asynchronous.  Here we examine
the possibilities.

Let us split the
planet into an ``atmosphere'', a part of small mass for which
thermal effects are significant, and an ``interior'' encompassing most
of the mass which has minimal horizontal thermal contrasts.  Suppose
(since $\tau_{\rm syn}$ is short) that the system has reached steady
state.  Consider two cases, depending
on the physical mechanisms that
determine the gravitational torque on the atmosphere.

\begin{figure}[ht]
\begin{center}
\noindent\resizebox{\hsize}{!}{\includegraphics{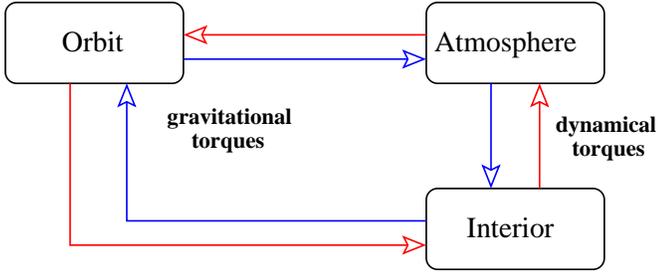}}
\caption{%
Angular momentum flow between orbit, interior, and atmosphere
for a \hotjup\s in steady state.  Arrows indicate flow of
prograde angular momentum (i.e., that with the same sign as the
orbital angular momentum) for two cases: 
{\it Anticlockwise:} Gravitational torque on atmosphere is retrograde (i.e.,
adds westward angular momentum to atmosphere). For torque balance, the 
gravitational torque on the interior must be prograde (i.e., eastward).
These gravitational torques must be balanced by fluid-dynamical torques
that transport retrograde angular momentum from atmosphere to interior. 
{\it Clockwise:} Gravitational torque
on atmosphere is prograde, implying a retrograde torque on the
interior and downward transport of
prograde angular momentum from atmosphere to interior.  Atmosphere
will superrotate if gravitational torques push atmosphere away from 
synchronous (as on Venus). It will subrotate if gravitational torques 
synchronize the atmosphere (e.g., gravity-wave resonance; cf. Lubow
et al. (1997)).}
\label{fig:boxes}
\end{center}
\end{figure}

The first possibility is that the gravitational torque on the atmosphere
pushes the atmosphere away from synchronous rotation (i.e., it
increases the magnitude of the atmosphere's angular momentum measured
in the synchronously-rotating reference frame)
as has been hypothesized for Venus (Ingersoll and Dobrovolskis
\cite{Dob78}, Gold and Soter \cite{Gol69}).    
To balance the torque on the atmosphere, the interior
must have a net angular momentum of the same sign as the atmosphere
(so that both either super- or subrotate). In Fig.~\ref{fig:boxes},
superrotation then corresponds to a clockwise flux of angular momentum,
whereas subrotation corresponds to the anticlockwise scenario.

The second possibility is that the gravitational torque on the atmosphere
tends to synchronize the atmosphere.  This possibility may be relevant
because, on a gas-giant planet, it is unclear that the high-temperature 
and high-pressure regions would be $90^\circ$ out of phase, as is
expected on a terrestrial planet.  If the interior responds
sufficiently to atmospheric perturbations to keep the deep
isobars independent
of surface meteorology (an assumption that seems to work well in modeling
Jupiter's cloud-layer dynamics), then high-pressure and
high-temperature regions will
be {\it in} phase, which would sweep the high-mass regions downwind
and lead to a torque
that synchronizes the atmosphere.  Furthermore, if a resonance occurs
between the
tidal frequency and the atmosphere's wave-oscillation frequency, the
resulting gravitational
torques also act to synchronize the atmosphere (Lubow et al. 1997).
In this case, the interior and atmosphere have net angular momenta of opposite
signs. Depending on the sign of the atmosphere/interior momentum flux,
this situation would correspond in Fig.~\ref{fig:boxes} to either the
clockwise or anticlockwise circulation of angular momentum cases.

In both cases above, the gravitational torque on the atmosphere arises 
because of spatial density variations associated with surface meteorology,
which are probably confined to pressures less than few hundred bars.
Therefore, this torque acts on only a small fraction of the planet's
mass.  In contrast, the gravitational torque on the gravitational
tidal bulge affects a much larger fraction of the planet's mass.  It
is not necessarily true that the torque on the atmosphere is negligible
in comparison with the torque on the interior, however, because the
gravitational tidal bulge is only slightly out of phase with the
line-of-sight to the star (e.g., the angle is $10^{-5}$ radians for
Jupiter). In contrast, the angle between the 
meteorologically-induced density variations and the line of sight
to the star could be up to a radian.  Detailed calculations
of torque magnitudes would be poorly constrained, however, and will be
left for the future.

Several mechanisms may act to transfer angular momentum between the atmosphere
and interior.  Kelvin-Helmholtz shear instabilities, if present, would smooth
interior-atmosphere differential rotation.  On the other hand, waves transport
angular momentum and, because they act nonlocally, often {\it induce}
differential rotation rather than removing it.  The quasi-biennial oscillation
on Earth is a classic example (see, e.g., Andrews et al. 1987, Chapter 8).
Atmospheric tides (large-scale waves forced by the solar heating) are another
example of such a wave; on Venus, tides play a key role in increasing the
rotation rate of the cloud-level winds.
Finally, vertical advection may cause angular momentum
exchange between the atmosphere and the interior of \hotjups, 
because the angular momentum
of air in updrafts and downdrafts need not be the same. 

A simple estimate illustrates the extent of nonsynchronous rotation
possible in the interior.  Suppose that the globally-averaged flux
of absorbed starlight is $\fabs$, which is of order $10^8\ergscm$ for
\hotjups\s near 0.05 AU, and that the globally-averaged flux of
kinetic energy transported from the atmosphere to the interior 
is $\eta \fabs$,
where $\eta$ is small and dimensionless.  If this kinetic energy
flux is balanced by dissipation in the interior
with a spindown timescale of $\tau_{\rm syn}$, then the deviation of
the rotation frequency from synchronous is
\begin{equation}
\omega - \omega_s = \left({4 \pi \eta \fabs \tau_{\rm syn}\over
k^2 M}\right)^{1/2}
\end{equation}
This estimate is an upper limit for the {\it globally averaged}
asynchronous rotation rate because it assumes that kinetic
energy transported into the interior has angular momentum of a single
sign.  The global-average asynchronous rotation could be even lower if 
angular momenta transported downward in different regions have
opposite signs.
  
The spindown time is uncertain and depends
on the planet's $Q$ according to Eq.~(1).  To date, only one estimate
for the $Q$ of a \hotjup\s exists (Lubow et al. \cite{Lub97}), which
suggests $Q\sim 100$, at least during the early stages of spindown.
Orbital constraints on the natural satellites
of the giant planets imply that, at periods of a few days,
the tidal $Q$ of Uranus, Neptune, and 
Jupiter are of order $\sim10^5$ (Tittemore and Wisdom \cite{Tit89},
Banfield and Murray \cite{Ban92}, Peale \cite{Pea99}), which suggests
a spindown time of a few$\,\times10^5$ years (Eq.~1).  Although
the mechanisms that determine these planets' $Q$ values 
remain uncertain, likely possibilities include friction in the
solid inner core (Dermott \cite{Der79}) or overturning of tidally-forced 
waves in the fluid envelope (Ioannou and Lindzen \cite{Ioa93},
Houben et al. \cite{Hou01}).  Both mechanisms are possible for
\hotjups, and the wave-dissipative mechanism
should be more effective for \hotjups\s than
for Jupiter because of the thicker stable (radiative) layer in the former
(Houben et al. \cite{Hou01}).  These calculations suggest that
spindown times of $\sim10^5$--$10^6$ years are likely, although
larger values cannot be ruled out.

As discussed in Section 4, experience with planets in our solar system
suggests that atmospheric kinetic energy is generated at a
flux of $10^{-2} \fabs$, and if all of this energy enters the
interior, then $\eta \sim 10^{-2}$.  
Using a spindown time of $3\times10^5\,$years
implies that $\omega - \omega_s \sim 2\times 10^{-5}\sec^{-1}$,
which is comparable to the synchronous rotation frequency. The implied
winds in the interior are then of order $\sim2000\m\sec^{-1}$.  Even
if $\eta$ is only
$10^{-4}$, the interior's winds would be $200\m\sec^{-1}$.  The implication
is that the interior's spin could be asynchronous by up to a factor of two,
depending on the efficiency of energy and momentum transport into the 
interior.  If spindown times of $\sim10^6$ years turn out to be
serious under- or overestimates, then greater or lesser asynchronous 
rotation would occur, respectively.

\section{Atmospheric Circulation: Possible regimes and influence on
evolution}
\label{sec:dynamics}


\subsection{Basic Parameter Regime}

Rotation plays a central role in the atmospheric dynamics of
\hotjups, and HD209458b in particular.  The ratio of nonlinear
advective accelerations
to Coriolis accelerations in the horizontal momentum equation is
$u f^{-1} L^{-1}$ (called the Rossby number), where
$u$ is the mean horizontal wind speed, $f=2\Omega \sin\phi$ is
the ``Coriolis parameter'' (e.g. Holton \cite{Hol92}, pp. 39-40), $\Omega$
is the rotational angular velocity, $\phi$ is latitude, and $L$
is a characteristic length scale.  Rossby numbers of 0.03--0.3
are expected for winds of planetary-scale and speeds ranging
from 100--$1000\m\sec^{-1}$. In Section 3 we showed that modest 
asynchronous rotation may occur, in which case the Rossby number could
differ from this estimate by a factor up to $\sim 2$.
These estimates suggest that nonlinear advective terms are small compared
to the Coriolis accelerations, which must then 
balance with the pressure-gradient accelerations.
(The Rossby number could be $\sim1$ if the winds reach 
2000--$3000\m\sec^{-1}$, which we show below is probably the maximum
allowable wind speed.)

\begin{table*}[htb]
\caption{Rhines' and deformation lengths for giant planets}
\label{tab:param}
\begin{tabular}{lllllll}
\hline\hline

\hline
  &$u$  &$\radius$  &$\Omega$ &$L_{\beta}$  &$L_D$ &Jet width     \\
  &$(\m\sec^{-1})$ &($10^7\m$) &($10^{-4}\sec^{-1}$) &($10^7\m$) &($10^7\m$)
 &($10^7\m$) \\
\hline
Jupiter &50  &7.1 &1.74 &1.0 &$0.2$  &$\sim1$\\
Saturn &200  &6.0 &1.6  &1.9 &$0.2$   &$\sim2$ \\
Uranus &300  &2.6 &1.0  &2.0 &$0.2$   &$\sim2$\\
Netpune&300 &2.5 &1.1 &1.9   &$0.2$   &$\sim2$\\
HD209458b &? &10 &$\sim 0.2$ &$15(u/1\km\sec^{-1})^{1/2}$ &$\sim4$ &?\\

\hline
\end{tabular}

{\it Note.} $L_D$ calculated at the tropopause using $N=0.01\sec^{-1}$
for Saturn, Uranus, and Neptune, and $N=0.02\sec^{-1}$ for Jupiter.

\end{table*}

The zonality of the flow can be characterized by the Rhines'
wavenumber, $k_{\beta} \sim
(\beta/u)^{1/2}$, where $\beta$ is the derivative of $f$ with
northward distance (Rhines 1975).  The
half-wavelength implied by this wavenumber, called the Rhines'
scale $L_{\beta}$, provides a
reasonable estimate for the jet widths on all four outer planets 
in our solar system (Cho and Polvani 1996; see Table 1).  
For HD209458b, the Rhines' scale is $\sim1.5\times10^{10}
(u/1000\m\sec^{-1})^{1/2} \cm$, which exceeds the planetary radius if
the wind speed exceeds about $400\m\sec^{-1}$.

Another measure of horizontal structure is
the Rossby deformation radius (Gill 1982, p.\ 205),
$L_D \sim N H/f$, where $H$ is the scale height and $N$ is the Br\"unt-Vaisala
frequency (i.e., the oscillation frequency for a vertically displaced
air parcel; Holton \cite{Hol92}, p.\ 54).  
At pressures of a few bars, the temperature profiles
calculated for irradiated extrasolar giant planets by Goukenlouque
et al. (2000) suggest $N\sim0.0015\sec^{-1}$.  With a scale height
of $700\km$,  the resulting deformation radius is 40,000 km.  In contrast,
the deformation radii near the tropopause of the Jupiter, Saturn,
Uranus, and Neptune are of order $2000\km$ (Table 1).

The estimated Rhines' scale (for winds of $>500\m\sec^{-1}$, which we
show later are plausible speeds) and deformation radius of \hotjups\s
are similar to the planetary radius, and they are
a larger fraction of the planetary radius than is the case for
Jupiter, Saturn, Uranus, and Neptune (Table 1).  This fact suggests that
eddies may grow to hemispheric scale in the atmospheres of
\hotjups\s and that, compared with the giant planets in our
solar system, the general circulation hot Jupiters may be more
global in character. Unless the winds are extremely weak, \hotjups\s
are unlikely to have $>10$ jets as do Jupiter and Saturn.

An upper limit on the atmospheric wind speed can be derived from
shear-instability considerations.  We assume that no zonal winds are
present ($u(P_0)=0$) in the convective core, a consequence of
synchronization by tidal friction.
The build-up of winds at higher altitudes in the radiative envelope
is suppressed by Kelvin-Helmholtz instabilities
if the shear becomes too large. This occurs
when the Richardson number becomes smaller than 1/4 (cf. Chandrasekhar
\cite{Cha61}), i.e. when
\begin{equation}
Ri={N^2\over (du/dz)^2} < {1\over 4},
\label{eq:Ri}
\end{equation}
In the perfect gas approximation, assuming a uniform composition,
\begin{equation}
N^2={g\over H}(\gradad-\nabla_T).
\end{equation}
where $\gasconst$ is the
universal gas constant divided by the mean molecular mass, $c_p$ is
the specific heat, $H=\gasconst T/g$ is the pressure scale height,
$\gradad=\gasconst/c_p$ is the adiabatic gradient, 
$\nabla_T=d\ln T/d\ln p$, $T$ is temperature, and $p$ is pressure.
The hypothesis of uniform composition is
adequate despite hydrogen dissociation because the timescales involved
are generally much longer than the dissociation timescales.
It can be noted that a shear instability could appear at smaller
Richardson numbers in the presence of efficient radiative diffusion
(Zahn \cite{Zah92}; Maeder \cite{Mae95}). This possibility
will not be examined here.

The maximal wind speed at which Kelvin-Helmholtz instabilities occur
can then be derived by integration of Eq.~(\ref{eq:Ri}):
\begin{equation}
u_{\rm max}(P)\sim {1\over 2}\int_{P_0}^{P}
\left[{\cal R}T(\gradad-\nabla_T)\right]^{1/2} d\ln p,
\end{equation}
The value of $u_{\rm max}$ thus derived at $P\sim 1\,$bar is of the order of
$3000\rm\,m\,s^{-1}$, to be compared to the winds of
Jupiter, Saturn, Uranus and Neptune which reach
100--$500\rm\,m\,s^{-1}$ (Ingersoll et al. \cite{Ing95}).
Our estimate for \hotjups, assuming the winds are measured in the
synchronously-rotating frame, may be uncertain by a factor of $\sim2$
due to the possibility of a non-synchronously-rotating interior.
In comparison, the expected speed of sound at the tropopause of
HD209458b is $\sim2400\m\sec^{-1}$.

A characteristic timescale for zonal winds to redistribute temperature
variations over scales similar to the planetary radius $\radius$
then stems from $\tau_{\rm zonal}\wig{>}\radius/u_{\rm max}$.

The radiative heating timescale can be estimated by a ratio between the
thermal energy within a given layer and the layer's net radiated flux.
In the absence of dynamics, absorbed solar fluxes balance the
radiated flux, but dynamics perturbs the temperature profile away from
radiative equilibrium.  Suppose the radiative equilibrium temperature at
a particular location is $T_{\rm rad}$ and the actual temperature 
is $T_{\rm rad} + \Delta T$.
The net flux radiated towards outer space is then $4\sigma T_{\rm rad}^3
\Delta T$ and the radiative timescale is
\begin{equation}
\tau_{\rm rad}\sim \frac{P}{g}\frac{c_p}{4\sigma T^3},
\label{eq:t_rad}
\end{equation}
where $\sigma$ is the
Stefan-Boltzmann constant.

\begin{figure}[ht]
\begin{center}
\resizebox{\hsize}{!}{\includegraphics{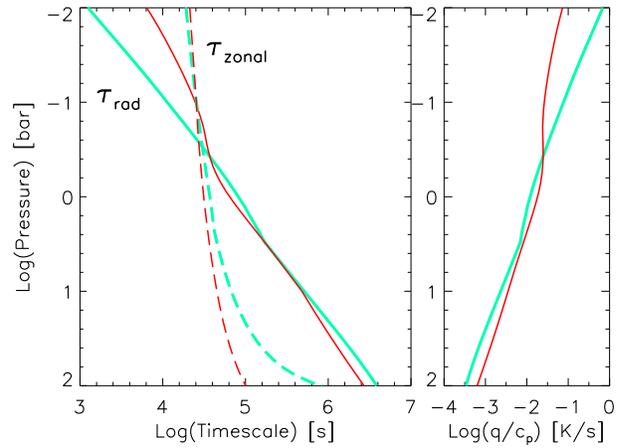}}
\caption{%
{\it Left:} Characteristic time scales as a function of pressure level.
$\tau_{\rm zonal}$ is the minimal horizontal advection time (dashed). $\tau_{\rm
rad}$ is the timescale necessary to cool/heat a layer of pressure $P$ and
temperature $T$ by radiation alone assuming the gas is optically thick 
(solid). (The plotted radiative timescale is an underestimate at pressures
less than about 0.3 bars because the atmosphere becomes optically thin 
at those pressures.)  For each case, the thin black and
thick grey lines correspond to the hot and cold models from Section 2
and Fig.\protect\ref{fig:profiles}.  {\it Right:} Approximate
cooling/heating rate as a function of pressure (see Eq.~\ref{eq:heat-rate}).}
\label{fig:timescales}
\end{center}
\end{figure}

Figure~\ref{fig:timescales} shows estimates of $\tau_{\rm zonal}$ and
$\tau_{\rm rad}$ for HD209458b
calculated using the temperature profiles from the
``hot'' (thin line) and ``cold'' (thick grey line) models from Section 2.
The zonal timescale is estimated by calculating the maximum wind speed that can
exist as a function of pressure given the static stability associated with
each model, while the radiative time is calculated using the
temperatures and heat capacities shown in Fig.~\ref{fig:profiles}.
At pressures exceeding $0.1$\,bar, radiation is
slower than the maximal advection by zonal winds, but by less than one
order of magnitude. The consequent day/night temperature difference $\Tdn$ to
be expected is:

\begin{equation}
{\Tdn \over \Delta T_{\rm rad}}\sim 1 - e^{-\tau_{\rm zonal}/\tau_{\rm rad}}.
\end{equation}
where $\Delta T_{\rm rad}$ is the day-night difference in radiative
equilibrium temperatures.
Rough estimates from Fig.~\ref{fig:timescales} suggest that 
$\tau_{\rm zonal}/\tau_{\rm rad} \sim0.3$ at 1 bar, implying that
$\Tdn/\Delta T_{\rm rad}\sim$0.3.   If $\Delta T_{\rm rad}=1000\K$, this
would imply day-night temperature differences of 300\,K at 1 bar.
Values of $\Tdn$ even closer to $\Delta T_{\rm rad}$ are likely given
the fact that slower winds will lead to an even more effective cooling on
the night side and heating on the day side.

The small radiative time scale implies that, for the day-night 
temperature difference to be negligible near the planet's photosphere,
atmospheric winds would have to be larger than the maximum winds for
the onset of shear instabilities.

\subsection{Nature of the circulation}

An understanding of the horizontal temperature difference and
mean wind speed is desirable.   It is furthermore of interest to
clarify the possible geometries the flow may take.  For \hotjups,
we envision that the dominant forcing is the large-scale
day-night heating contrast, with minimal role for moist convection.
This situation differs from that of Jupiter, where differential escape of the
intrinsic heat flux offsets the solar heating contrast and moist convection
plays a key role.  For \hotjups, such an offset between intrinsic and
stellar fluxes cannot occur, because the intrinsic flux is $10^4$
times less than 
the total flux. The fact that the day-night heating
contrast occurs at hemispheric scale --- and that the 
relevant dynamical length
scales for \hotjups\s are also hemispheric (Section 4.1) --- increases our 
confidence that
simple analysis, focusing on the hemispheric-scale circulation, can 
provide insight.

Because the Rossby number is small, the dominant balance in the horizontal
momentum equation at mid-latitudes 
is between the Coriolis force and the pressure-gradient force
(geostrophic balance). 
We adopt the primitive equations, which are the standard set of large-scale
dynamical equations in a stably-stratified planetary atmosphere.
When differentiated with pressure (which is used here as a vertical
coordinate), this balance leads to the well-known
thermal wind equation (e.g. Holton \cite{Hol92}, p.~75):

\begin{equation}
f{\partial {\bf v}\over\partial\ln p}= -\gasconst{\bf k\times
\nabla_H T}
\label{eq:thermal-wind}
\end{equation}
where ${\bf v}$ is the horizontal wind, $p$ is pressure,
$\nabla_H$ is the horizontal gradient, $T$ is temperature, and
${\bf k}$ is the unit upward vector.
Assuming the interior winds are small compared to the atmospheric winds,
this implies that, to order-of-magnitude,

\begin{equation}
|{\bf v}| \sim {\gasconst\over f \radius}\Thor \Delta\ln p
\label{eq:mom}
\end{equation}
where ${\bf |v|}$ and $\Thor$ are the characteristic
wind speed and horizontal temperature difference in the atmosphere and
$\Delta\ln p$ is the difference in log-pressures from bottom to
top of the atmosphere.  Non-sychronous rotation of the interior would
alter the relation by changing the value of $f$, but the uncertainty
from this source is probably a factor of two or less (see Eq.~(6) and
subsequent discussion).

The thermodynamic energy equation is, using pressure as a vertical 
coordinate (e.g. Holton 1992, p. 60),

\begin{equation}
{\partial T\over \partial t} + {\bf v}\cdot\nabla_{\rm H} T -
\omega {H^2 N^2\over \gasconst p} = {q\over c_p}
\label{eq:heat}
\end{equation}
where $t$ is time, $\omega = dp/dt$ is vertical velocity,
$q$ is the specific heating rate (erg${\rm\,g}^{-1}{\rm\,sec}^{-1}$),
and we have assumed an ideal gas.

A priori, it is unclear whether the radiative heating and cooling is
dominantly balanced by horizontal advection (second term on left of Eq.~(14))
or vertical advection (third term on the left).  To illustrate the
possibilities, we consider two endpoint scenarios corresponding to
the two limits.  In the first scenario, the radiation is balanced
purely by horizontal advection: zonal winds transport heat from
dayside to nightside, and meridional winds transport heat from
equator to pole.  In the second scenario, the radiation is balanced by
vertical advection (ascent on dayside, descent on nightside).

Consider the first scenario, where horizontal
advection dominates.  Generally, we expect ${\bf v}$ and $\nabla_{\rm H} T$
to point in different directions, so to order of magnitude 
their dot product
equals the product of their magnitudes.  (If the direction of
$\nabla_{\rm H} T$ is independent of height and no deep 
barotropic flow exists, then at mid-latitudes one could argue that winds
and horizontal pressure gradients are perpendicular to order $Ro$.
However, the existence of either asynchronous rotation or variations
in the orientations of $\nabla_{\rm H} T$ with height will imply
that winds and $\nabla_{\rm H} T$ are not perpendicular even if
$Ro<<1$.  Because asynchronous rotation and variation
in the directions of $\nabla_{\rm H} T$ with height are likely,
and because, in any case, ${\bf v}$ and $\nabla_{\rm H} T$ will not be 
perpendicular near the equator, we write the dot product as the
product of the magnitudes.) An order-of-magnitude form of the
energy equation is then

\begin{equation}
{{\bf |v|} \Thor\over\radius}\sim {q\over c_p}
\end{equation}

The solutions are

\begin{equation}
{\bf |v|}\sim\left({q\over c_p} {\gasconst\Delta\ln p \over f}\right)^{1/2} 
\label{eq:u}
\end{equation}

\begin{equation}
\Thor\sim\radius \left({q\over c_p} {f\over \gasconst \Delta\ln p}\right)^{1/2} 
\label{eq:deltat}
\end{equation}

A rough estimate of the heating rate, $q/c_p$, results
from the analysis in Section 4.1:

\begin{equation}
{q\over c_p} = {4\sigma T^3 \Delta T g\over p c_p}
\label{eq:heat-rate}
\end{equation}
where $\Delta T$ is the characteristic magnitude of the difference between
the actual and radiative equilibrium temperatures in the atmosphere.
The heating rate depends on the dynamics through $\Delta T$.  
We simply evaluate the heating
rate using Eq.~(\ref{eq:heat-rate}) with $\Delta T \approx T/2$. The
results are shown in Fig.~\ref{fig:timescales}.

The key difficulty in applying the equations to \hotjups\s
is the fact that $q/c_p$ depends on pressure, and 
$\Thor$ probably should too,
but Eqs.~(16) and (17) were derived assuming that $\Thor$ is constant.
We can still obtain rough estimates by inserting values of $q/c_p$
at several pressures.  At $\sim 50$--100 bars, where
$q/c_p\sim 10^{-4}$--$10^{-3}\K\sec^{-1}$, we obtain temperature
differences and wind speeds of 50--$150\K$ and 200--$600\m\sec^{-1}$.
At 1 bar, where $q/c_p$ reaches $10^{-2}\K\sec^{-1}$, the estimated
temperature contrast and wind speed is $\sim500\K$ and $\sim2000\m\sec^{-1}$.
The estimates all assume $f\approx3\times10^{-5}\sec^{-1}$,
$\gasconst=3500\J\kg\K^{-1}$, and $\Delta\ln p\approx 3$.

Later we show that the equations successfully
predict the mean wind speeds and temperature differences obtained in
numerical simulations of the circulation of HD209458b.  
This gives us confidence in the results.  

Now consider
the second scenario, where vertical advection (third term on the left of
Eq.~(14)) balances the radiative heating and cooling.
The magnitude of $\omega$ can be estimated from the continuity equation.
Purely geostrophic flow has zero horizontal divergence, so $\omega$ is
roughly

\begin{equation}
\omega \sim Ro {|{\bf v}| \over \radius}p
\end{equation}
Strictly speaking, this is an upper limit, because the nonlinear terms
comprising the numerator of the Rossby number contain some terms
(e.g., the centripetal acceleration) that are divergence-free.
Substituting this expression into the energy equation, using Eq.~(13), and
setting $Ro = |{\bf v}|/f \radius$ implies that

\begin{equation}
|{\bf v}|\sim {\radius\over N H} \left( {q\over c_p} f \gasconst \right)^{1/2}
\end{equation}
\begin{equation}
\Thor \sim {f \radius^2\over N H \Delta\ln p}\left({q\over c_p}
{f\over \gasconst}\right)^{1/2}
\end{equation}

Numerical estimates using $H=700\km$, $N\sim0.0015\sec^{-1}$, and 
$q/c_p=10^{-4}$--$10^{-3}\K\sec^{-1}$ (appropriate to 50--100 bars)
yield temperature differences and speeds of 80--$250\K$ and 
300--$900\m\sec^{-1}$, respectively.  Here we have used the same
values for $\radius$ and $f$ as before.  A heating rate of 
$10^{-2}\K\sec^{-1}$, appropriate at the 1 bar level, yields
$|{\bf v}|
\sim3000\m\sec^{-1}$ and $\Thor \sim 800\K$. 
These values are similar to those obtained when we
balanced heating solely against horizontal advection.  

We can estimate the vertical velocity under the assumption that
vertical advection balances the heating.  Expressing the vertical
velocity $w$ as the time derivative of an air parcel's altitude,
and using $w=-\omega/\rho g$, where $\rho$ is density,
yields

\begin{equation}
w\sim{q\over c_p}{\gasconst\over H N^2}
\label{eq:w}
\end{equation}
Using $q/c_p=10^{-2}\K\sec^{-1}$ and $H=700\km$ implies that
$w\sim 20\m\sec^{-1}$ near 1 bar.  
If this motion comprises the vertical branch of
an overturning circulation that extends vertically over a scale height
and horizontally over $\sim10^{10}$ cm (a planetary radius), the implied
horizontal speed required to satisfy continuity is $\sim3000\m\sec^{-1}$,
consistent with earlier estimates.

The numerical estimates, while rough, suggest that
winds could approach the upper limit of $\sim3000\m\sec^{-1}$
implied by the shear instability criterion.  This comparison suggests
that Kelvin-Helmholtz shear instabilities may play an important role
in the dynamics.

The likelihood of strong temperature contrasts can also be seen from 
energetic considerations. The differential
stellar heating produces available
potential energy (i.e., potential energy that can
be converted to kinetic energy through rearrangement of the fluid;
Peixoto and Oort 1992, pp.~365-370), and in steady-state, this
potential energy must be converted to kinetic energy at the rate it
is produced.  This requires
pressure gradients, which only exist in the presence of lateral 
thermal gradients.  A crude estimate suggests that the rate of 
change of the difference in gravitational potential $\Phi$ between
the dayside and nightside on isobars caused 
by the heating is
$\sim \gasconst (q/c_p) \Delta\ln p$.  The rate per mass at which potential
energy is converted to kinetic energy by pressure-gradient work 
is ${\bf v}\cdot\nabla_H\Phi$, which is approximately 
${\bf |v|} \gasconst\Thor\Delta\ln p/\radius$.  Equating the two
expressions 
suggests, crudely, that $\Thor\sim (q/c_p)(\radius/{\bf |v|})$,
which implies 
$\Thor\sim 500\K$ using the values of $q/c_p$ and ${\bf |v|}$
near 1 bar discussed earlier.


One could wonder whether a possible flow geometry consists of dayside
upwelling,
nightside downwelling, and simple acceleration of the flow from
dayside to nightside
elsewhere (leading to a flow symmetrical about the subsolar point).
The low-Rossby-number considerations described above argue against this
scenario.  A major component of the flow must be perpendicular to horizontal
pressure gradients. Direct acceleration of wind from dayside to nightside is
still possible near the equator, however.  This would essentially be an
equatorially-confined Walker-type circulation that, as described earlier,
would lie equatorward of the geostrophic flows that exist at more 
poleward latitudes.

\subsection{Numerical Simulations of the Circulation}

To better constrain the nature of the circulation, we performed
preliminary three-dimensional, fully-nonlinear numerical simulations
of the atmospheric circulation of HD209458b.  For the calculations, we used
the Explicit Planetary Isentropic Coordinate, or EPIC, model
(Dowling et al. 1998).  The model solves the primitive equations
in spherical geometry using finite-difference methods and
isentropic vertical coordinates.  The equations are valid in
stably-stratified atmospheres, and we solved the equations within
the radiative layer from 0.01 to 100 bars assuming the planet's
interior is in synchronous rotation with the 3.5-day orbital period.  
The radius, surface gravity,
and rotation rate of HD209458b were used ($10^8\m$, $10\m\sec^{-2}$,
and $\Omega = 2.1\times10^{-5}\sec^{-1}$, respectively).  

The
intense insolation was parameterized with a simple
Newtonian heating scheme, which relaxes the temperature toward
an assumed radiative-equilibrium temperature profile.  The
chosen radiative-equilibrium temperature profile was hottest
at the substellar point ($0^{\circ}$ latitude, $0^{\circ}$ longitude)
and decreased toward the nightside. 
At the substellar point, the profile's height-dependence was 
isothermal at 550 K
at pressures less than 0.03 bars and had
constant Brunt-Vaisala frequency of $0.003\sec^{-1}$ at pressures
exceeding 0.03 bars,
implying that the temperature increased with depth.  The nightside
radiative-equilibrium temperature profile was equal to the 
substellar profile minus $100\K$; it is this day-night difference
that drives all the dynamics in the simulation.  The
nightside profile was constant across the nightside, and the 
dayside profile varied as
\begin{equation}
T_{\rm dayside} = T_{\rm nightside} + \Delta T_{\rm rad} \cos\alpha
\end{equation}
where $T_{\rm dayside}$ is the dayside radiative-equilibrium 
temperature at a given latitude and longitude, $ T_{\rm nightside}$ 
is the nightside radiative-equilibrium temperature, $\Delta T_{\rm rad}=100\K$,
and $\alpha$ is the angle between local vertical and the line-of-sight
to the star.  For simplicity, the timescale over which the temperature
relaxes to the radiative-equilibrium temperature was assumed constant
with depth with a value of $3\times10^5\sec$.  This is equal to
the expected radiative timescale at a pressure of about 5 bars (Fig. 4,
left).

The Newtonian heating scheme described above is, of course, a
simplification.  The day-night difference in radiative-equilibrium
temperature, 100 K, is smaller than the expected value. Furthermore,
the radiative timescale is too long in the upper troposphere
($\sim0.1$--1\,bar) and too short at deeper pressures
of $\sim10$--100 bars.  Nevertheless, the net column-integrated heating
per area produced by the scheme ($\W\m^{-2}$) is similar to that
expected to occur in the deep troposphere of HD209458b and other 
\hotjups, and the
simulation provides insight into the circulation patterns that
can be expected.

The temperature used as the initial condition was isothermal at 500 K
at pressures less than 0.03 bars and had a constant Brunt-Vaisala
frequency of $0.003\sec^{-1}$ at pressures exceeding 0.03 bars.  
There were no initial winds.  The simulations were performed with
a horizontal resolution of $64\times32$ with 10 layers
evenly spaced in log-pressure.

Figure 5--8 show the results of such a simulation.  Despite the
motionless initial condition, winds rapidly
develop in response to the day-night heating contrast, reaching
an approximate steady state after $\sim400$ Earth days.
Snapshots at 42 and 466 days are shown in Figs. 5 and 6, respectively.
In these figures, each panel shows pressure on an isentrope 
(greyscale) and winds (vectors) for three of the model layers 
corresponding to mean pressures of roughly 0.4, 6, and 100 bars 
(top to bottom, respectively).  The greyscale is such that,
on an isobar, light regions are hot and dark regions are cold.

\begin{figure}[htb]
\begin{center}
\resizebox{\hsize}{!}{\includegraphics[angle=0]{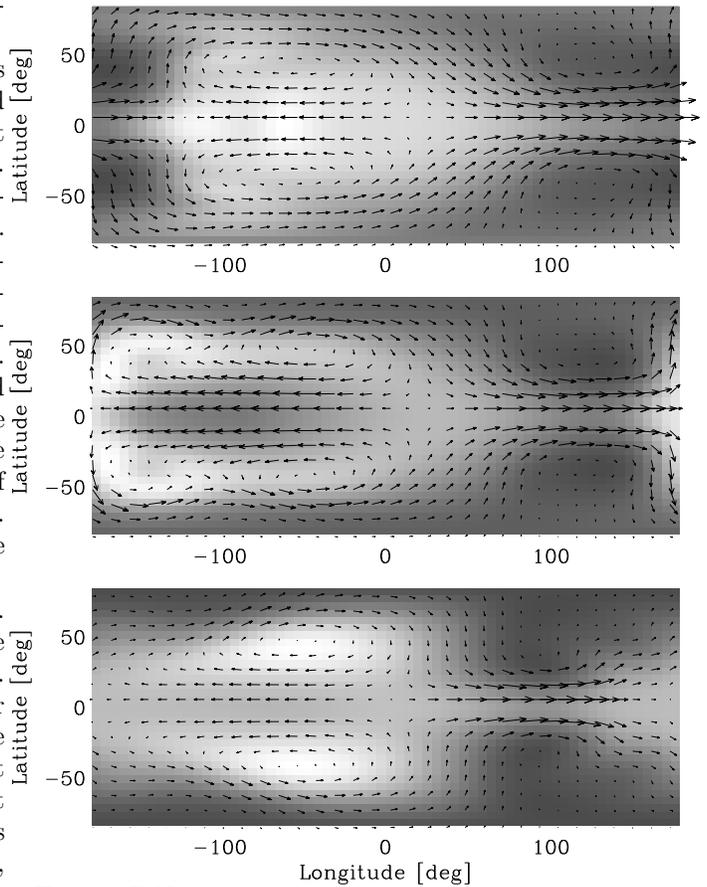}}
\caption
{EPIC simulation of atmospheric circulation on HD209458b after
42 days.  Panels depict pressure on an isentrope (greyscale) and winds 
(vectors) for three model levels with mean pressures near
0.4, 6, and 100 bars (top to bottom, respectively).  From
top to bottom, the maximum wind speeds are 937, 688, and
$224\m\sec^{-1}$, respectively, and the greyscales span (from
dark to white) 0.31--0.49 bars, 5.6--7.8 bars, and 92--113 bars,
respectively.  Substellar point is at $0^{\circ}$ latitude,
$0^{\circ}$ longitude.
}
\label{fig:simulation1}
\end{center}
\end{figure}

\begin{figure}[htb]
\begin{center}
\resizebox{\hsize}{!}{\includegraphics[angle=0]{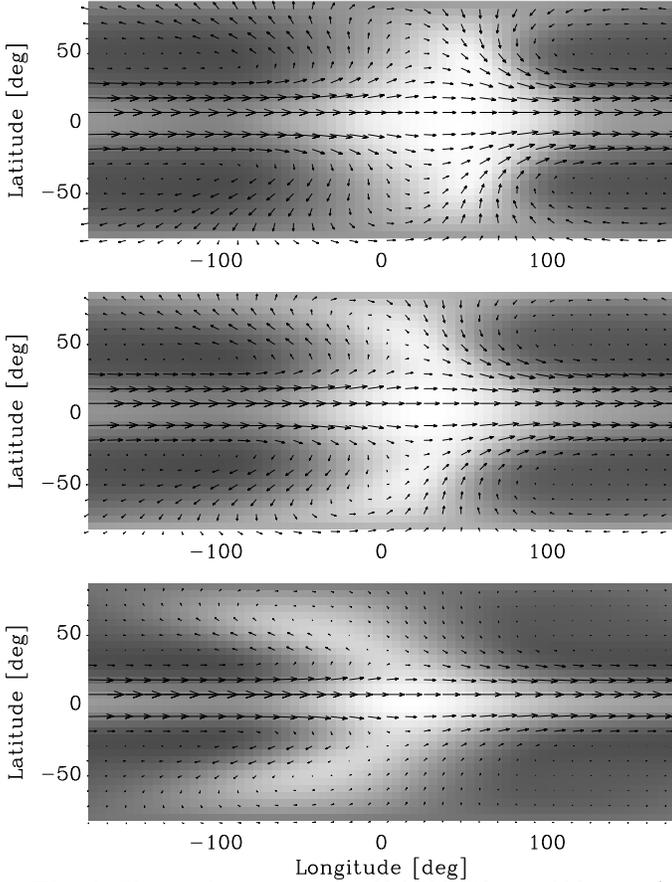}}
\caption
{Simulation results for HD209458b at 466 days (after a steady
state has been reached).   As in Fig. 5, panels depict pressure
on isentropes (greyscale) and winds (vectors) for three model levels 
with mean pressures near 0.4, 6, and 100 bars (top to bottom, respectively).  
From top to bottom, the maximum wind speeds are 1541, 1223, and
$598\m\sec^{-1}$, respectively, and the greyscales span (from
dark to white) 0.32--0.51 bars, 5.6--8.1 bars, and 90--127 bars,
respectively.  Substellar point is at $0^{\circ}$ latitude,
$0^{\circ}$ longitude.
}
\label{fig:simulation2}
\end{center}
\end{figure}

\begin{figure}[htb]
\begin{center}
\resizebox{\hsize}{!}{\includegraphics[angle=0]{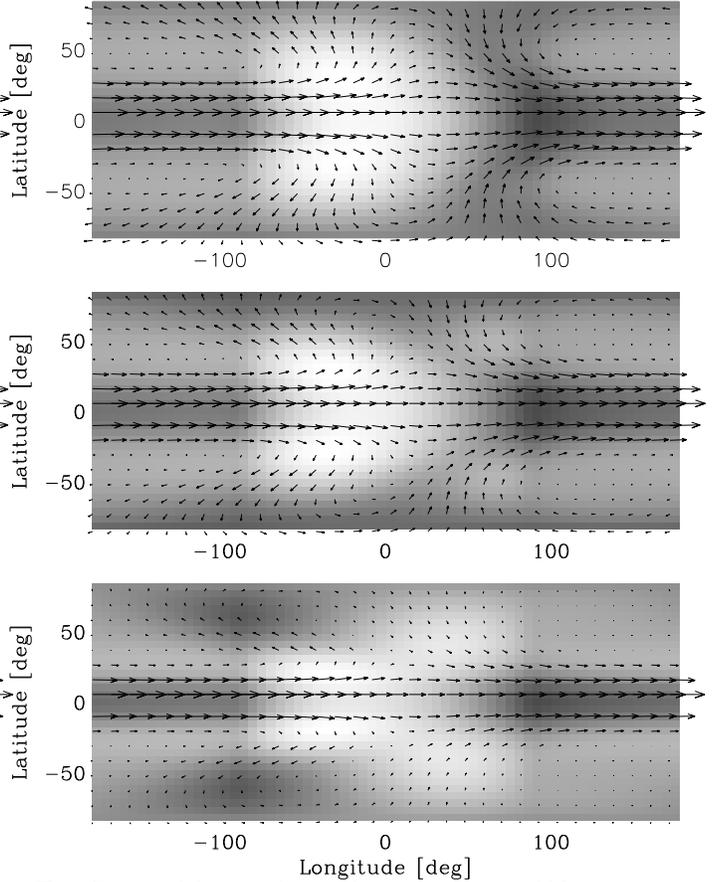}}
\caption
{Additional simulation results at 466 days.  Arrows are identical
to those in Fig. 6, but here greyscale is vertical velocity
$d\theta/dt$, where $\theta$ is potential temperature.  Light
regions are heating (i.e., $\theta$ is increasing) and
dark regions are cooling ($\theta$ is decreasing).  From
top to bottom, the greyscales span (from dark to white)
$-0.003$ to $0.003\K\sec^{-1}$, $-0.002$ to $0.002\K\sec^{-1}$,
and $-0.001$ to $0.001\K\sec^{-1}$, respectively.  The 
lowermost panel separates the convective interior from the
radiative region, and the implication is that mass exchange
can happen across this interface.
}
\label{fig:simulation3}
\end{center}
\end{figure}

The simulation exhibits several interesting features.
First, peak winds exceed $1\km\sec^{-1}$, but despite these
winds, a horizontal temperature contrast is maintained.
The dayside (longitudes $-90^{\circ}$ to
$90^{\circ}$) is on average hotter than the nightside, but
dynamics distorts the temperature pattern in a 
complicated manner.  Second, an equatorial jet
develops that contains most of the kinetic energy.  
The jet initially exhibits both eastward and westward 
branches (Fig. 5) but eventually becomes only eastward
(Fig. 6), and extends from $-30^{\circ}$ to $30^{\circ}$
in latitude.  Third, away from the equator, winds
develop that tend to skirt parallel to the temperature
contours.  This is an indication that geostrophic balance 
holds.  (Because the thermal structure is independent of 
height to zeroeth order and we have assumed no deep barotropic
flow, horizontal pressure and 
temperature gradients are parallel.)  Nevertheless, winds
are able to cross isotherms near the equator, and this is
important in setting the day-night temperature contrast.
As expected from the order-of-magnitude arguments in Section 4.1,
the jets and gyres that exist are broad in scale,
with a characteristic width of the planetary radius.

In the simulation,
the day-night temperature difference (measured on isobars)
is about $50\K$. This value depends on the adopted heating rate,
which was chosen to be appropriate to the region where the
pressure is tens of bars.  Simulations
that accurately predict the day-night temperature difference
at 1 bar will require a more realistic heating-rate scheme;
we will present such simulations in a future paper.

The temperature patterns in Figs. 5 and 6 show
that Earth-based infrared measurements can shed light
on the circulation of \hotjups.  In Fig. 6, 
the superrotating equatorial jet blows the high-temperature region
downwind.  The highest-temperature region is thus
{\it not} at the substellar point but lies eastward
by about $60^{\circ}$ in longitude.  The maximum
and minimum temperatures would thus face Earth 
{\it before} the transit of the planet behind
and in front of the star, respectively.  On the
other hand, if a broad westward jet existed instead,
the maximum and minimum temperatures would face Earth
{\it after} the transits.  Therefore, an infrared
lightcurve of the planet throughout its orbital cycle
would help determine the direction and strength of
the atmospheric winds.

In the simulation, the intense heating and cooling 
causes air to change
entropy and leads to vertical motion, as shown in
Fig. 7. Light regions (Fig. 7) indicate heating, which
causes ascent, and dark regions indicate
cooling, which causes descent.  Air is thus
exchanged between model layers.  This exchange also
occurs across the model's lowermost isentrope, which
separates the radiative layer from the convective
interior (Fig. 7, bottom).  Because upgoing and 
downgoing air generally have different kinetic energies and 
momenta, energy and momentum can thus be exchanged 
between the atmosphere and interior.

Figure 8 indicates how the steady state is achieved.
The build-up of the mass-weighted mean speed (top panel)
involves two timescales. Over the first 10 days,
the mean winds accelerate to 
$110\m\sec^{-1}$, and the peak speeds approach $1\km\sec^{-1}$.
This is the timescale for pressure gradients to accelerate 
the winds and force balances to be established.  The flow then 
undergoes an additional, slower ($\sim300$ day) increase in 
mean speed to $300\m\sec^{-1}$, with peak speeds of 
$\sim1.5\km\sec^{-1}$.  This timescale is that for exchange
of momentum with the interior, across the model's lowermost
isentrope, to reach steady state, as shown in Fig. 8, middle.

In the simulation, kinetic energy is transported from the
atmosphere into the interior at a rate that reaches $2500\W\m^{-2}$
(Fig. 8, bottom).  This downward transport of kinetic energy 
is $\sim 1\%$ of the absorbed stellar flux and
is great enough to affect the radius of HD209458b,
as shown in Paper I.  Although the simulation described here
does not determine the kinetic energy's fate once it reaches
the convective interior, we expect that
tidal friction, Kelvin-Helmholtz instabilities, or other
processes could convert these winds to thermal energy.  The exact
kinetic energy flux will depend on whether a deep barotropic
flow exists. Furthermore, potential and thermal energy are also transported 
through the boundary, and pressure work is done across it.
Our aim here
is not to present detailed diagnostics of the energetics,
but simply to point out that energy fluxes that are large enough to 
be important can occur.  We are currently conducting more detailed
simulations to determine the sensitivity of the simulations
to a deep barotropic flow, and we will present the
detailed energetics of these simulations in a future paper.

The evolution of  angular momentum (Fig. 8, middle)
helps explain why the circulation changes from Fig.
5 to 6.  After 42 days (Fig. 5), eastward midlatitude winds
have already developed that balance the negative 
equator-to-pole temperature gradient.  But at this 
time, the atmosphere's angular momentum is still nearly
zero relative to the synchronously-rotating state,
so angular momentum balance requires westward winds
along some parts of the equator.  (Interestingly, in these 
regions, the temperature
{\it increases} with latitude, as expected from geostrophy.)
The circulation involves upwelling on the dayside and
equatorial flow --- both east and west --- to the nightside, 
where downwelling occurs.  This circulation is similar
in some respects to the Walker circulation in Earth's
atmosphere.  After 466 days, however, the atmosphere
has gained enough angular momentum that the equatorial
jet is fully superrotating.  Some westward
winds exist at high latitudes, but they are weak enough
that the net angular momentum is still eastward.

Application of the arguments from Section 4.2
provides a consistency check.
The mean heating rate in the simulations is about
$1.7\times10^{-4}\K\sec^{-1}$.  Inserting this value
into Eqs.~(16--17), we obtain $\Thor = 70\K$
and $|{\bf v}| = 240\m\sec^{-1}$.  These compare well with
the day-night temperature difference and mass-weighted
mean speed of $50\K$ and $300\m\sec^{-1}$ that are obtained
in the simulation. (For the estimate, we used 
$f=3\times10^{-5}\sec^{-1}$, $\gasconst = 3500\J\kg^{-1}\K^{-1}$,
$\radius=10^8\m$, and $\Delta p=3$.) If Eqs.~(20--21) are adopted instead,
using values of $N=0.003\sec^{-1}$  and $H=300\km$, we
obtain temperature differences and speeds of $130\K$ and 
$470\m\sec^{-1}$.  Compared to the simulation, these estimates
are too high by a factor of $\sim2$, but are still of the
correct order of magnitude.  An important point is that
the {\it maximum} speed in the simulation substantially
exceeds that from the simple estimates; the estimates are
most relevant for the mean speed.

\begin{figure}[htb]
\begin{center}
\resizebox{\hsize}{!}{\hspace{0.8cm}\includegraphics[angle=0]{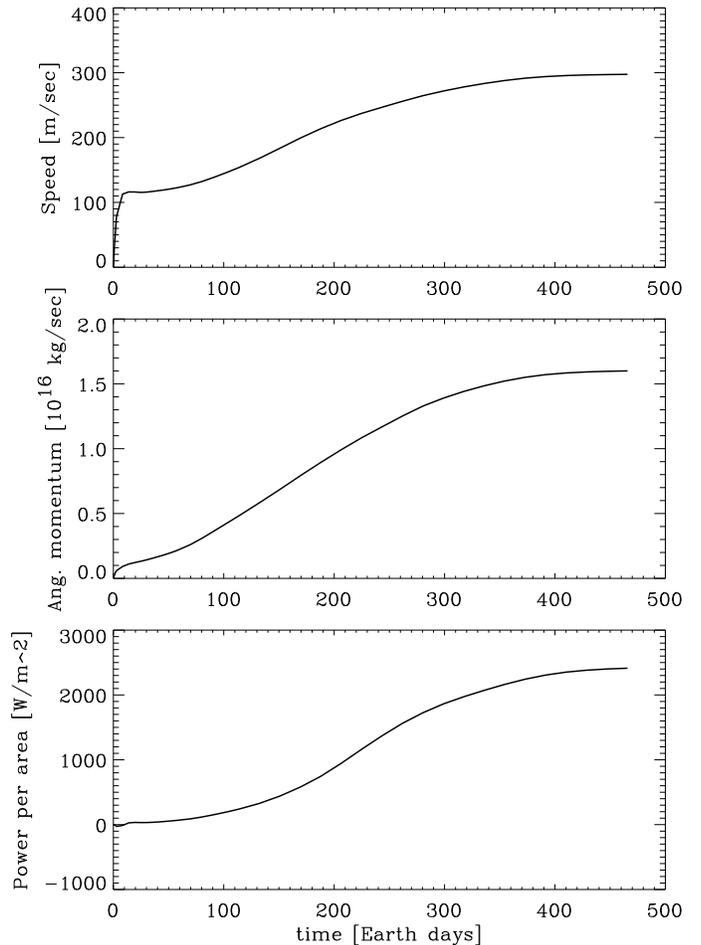}}
\caption
{ 
{\it Top:} Mass-weighted mean wind speed within the model domain
($\sim0.01$--100 bar) during the simulation of HD209458b shown in
Figs. 5--7. {\it Middle:} Globally-averaged zonal angular momentum
per area minus that of the synchronously-rotating state.
(That is, $\radius A^{-1} \int u\cos\phi\,dm$, where $A$ is
the planet's area, $u$ is eastward speed, $\phi$ is latitude, and $dm$ 
is an atmospheric mass element.) {\it Bottom:} Globally-averaged
flux of kinetic energy across the model's bottom isentrope
(at $\sim100$ bars), which is the interface between the radiative
layer and the convective interior in this model.
}
\label{fig:simulation4}
\end{center}
\end{figure}

\subsection{Vertical motion and clouds}

As shown in the appendix, hot and cold regions on the planet may
have distinct chemical equilibrium compositions. 
Advection of air between these regions plays a key role for the
(disequilibrium) chemistry in the atmosphere. Cloud formation
critically depends on whether vertical motions dominate
over horizontal motions. This in turns affects the albedo and depth to
which stellar light is absorbed.

Clouds form when the temperature of moving air parcels
decreases enough for the vapor pressure of condensible vapors to become
supersaturated.  If dayside heating and nightside cooling
are balanced only by horizontal advection,  air flowing from dayside
to nightside remains at constant pressure but decreases in temperature
(hence entropy) because of the radiative cooling.   Nightside condensation
may occur, and particle fallout limits the total mass of the trace
constituent (vapor plus condensates) in this air.  As the air flows onto the
dayside, its temperature increases and
any existing clouds will sublimate.  This scenario therefore
implies that the dayside will be cloud-free.

In an alternate scenario, dayside heating and nightside cooling are
balanced by vertical advection, at least near the sub- and antisolar
points.   Because entropy increases with altitude in a
stably-stratified atmosphere, this scenario implies that ascent occurs
on the dayside and descent on the nightside.  Clouds may therefore
form on the dayside, increasing the albedo and decreasing the
pressure at which stellar light is absorbed.

In reality, both horizontal and vertical advection are important, and the
real issue is to determine the relative contribution. We are pursuing
more detailed numerical simulations to address this issue.

\section{Conclusions}

We examined the structure and dynamics of the atmospheres of \hotjups\s
and concluded that 
strong day-night temperature contrasts ($300\K$ or more)
are likely to occur near the level where radiation is emitted to space
($\sim 1$ bar). These temperature variations should drive a
rapid circulation with peak winds of $1\km\sec^{-1}$ or more.
Force-balance arguments suggest that the mean midlatitude winds are eastward,
but the equatorial winds could blow either east or west.
Depending on the dynamical
regime, the cloud coverage and consequently the radiative absorption
of the incoming stellar flux will be very different.

Preliminary numerical simulations show that, while the dayside
is generally hotter than the nightside, the specific temperature
distribution depends on the dynamics. In our simulations,
a broad superrotating jet develops that
sweeps the high-temperature regions downwind by about 1 radian.
This implies that the greatest infrared flux would reach Earth
$\sim10$--15 hours before the occultation of the planet behind the 
star.  On the other hand, if a subrotating jet existed, the
greatest infrared flux would reach Earth {\it after} the occultation.
Measurements of the infrared lightcurve of \hotjups\s will 
therefore constrain the direction and magnitude of the wind.

The simulations also produce a downward flux of kinetic energy
across the $\sim100$ bar surface equal to $\sim1\%$ of the absorbed
stellar flux.  Although the simulation did not explicitly include
the convective interior, we surmise that a substantial fraction
of this kinetic energy would be converted to thermal energy by
Kelvin-Helmholtz instabilities and other processes.  As discussed
in Paper I, ``standard'' evolution models explain HD209458b's radius only
when the atmosphere is assumed to be unrealistically hot, but addition
of an internal energy source allows a more realistic model to match the observed
radius. The downward transport and subsequent dissipation of kinetic
energy described here is a promising candidate.  Bodenheimer et al.
(2001) suggested an alternative --- that the internal heating could
be provided by tidal circularization of an initially eccentric
orbit.  The difficulty, as Bodenheimer et al. were careful to point out,
is that the tidal heating is a transient process in the absence
of a detected close, massive companion capable of exciting the planet's
eccentricity.  In contrast, the mechanism proposed here can last
throughout the star's lifetime. Nevertheless, to fully determine
the feasibility of the mechanism, more detailed
numerical simulations are required (in particular, to test the
dependence of the energetics on the possible existence of winds in
the interior).  We will present such simulations in a future paper.

Upcoming observations are likely to provide key information
within the decade.  Several spacecraft
missions (either proposed or accepted) and dedicated ground programs
will observe extrasolar planets.  Measurement of starlight reflected
from these planets may allow the albedo to be estimated.
Because the star-planet-Earth angle changes
throughout the planet's orbit, crude information on the scattering properties of
the atmosphere (e.g., isotropic versus forward scattering) may be obtainable.
Asymmetries in the reflected flux as the planet approaches and recedes from
the transit could give information on the differences of albedo near the
leading and trailing terminators, which would help constrain the dynamics.
Finally,  transit observations of \hotjups\s using high
resolution spectroscopy should in the near future yield
constraints on the atmospheric temperature, cloud/haze abundance,
and winds (Brown \cite{Bro01};
see also Seager \& Sasselov \cite{SS00}; Hubbard et
al. \cite{Hub01}). If
these measurements are possible during the ingress and egress,
i.e., the phases during which the planets enters and leaves the stellar
limb, respectively, asymmetries of the planetary signal should be
expected and would indicate zonal heat advection at the
terminator. The duration of these
phases being limited to less than 10 minutes, it is not clear that
this effect is observable with current
instruments.

\begin{acknowledgements}
We wish to thank F. Allard, P. Bodenheimer,
H. Houben, S. Peale, 
D. Saumon, D.J. Stevenson, R.E. Young and K. Zahnle for a variety of
useful contributions,
and T. Barman for sharing results in advance of publication.
This research was supported by the French {\it Programme National de
Plan\'etologie}, Institute of Theoretical Physics (NSF PH94-07194), and
National Research Council of the United States.
\end{acknowledgements}

\appendix

\section{Day-night temperature variations and chemical composition}

Here we explore the implications of day-night temperature
variations on the abundances of a condensing species. Suppose
that on the day side, a
given condensate (e.g. iron) condenses at a pressure level
$\psday$ and a corresponding temperature $\tday$. If the temperature
on the night side $\tnight$ is smaller, at what pressure $\psnight$
will condensation take place? 

Assuming an ideal gas, we write the Clausius-Clapeyron
equation as 
\begin{equation}
{d\ln p\over d\ln T}=\beta
\label{eq:clausius}
\end{equation}
where $p$ is the partial pressure of saturation of the condensing
species and $\beta=L/\gasconst T$ is the ratio of the latent
heat of condensation $L$ to the thermal energy $\gasconst T$. For most
condensing species of significance here, $\beta\approx 10$. We
furthermore assume that $\beta$ is independent of $T$ and $P$, which
introduces only a slight error in our estimates. 

By definition, on the day side, the saturation abundance of the
condensing species, $x=p/P$ is maximal and equal to $x^\star$ at
$P=\psday$. On the other hand, the night side temperature is
lower and the abundance becomes:
\begin{equation}
\ln x(\psday)=\ln x^\star - \beta \ln (\tday/ \tnight).
\label{eq:xnight}
\end{equation}
In order to reach condensation, i.e. $x=x^\star$, one has to penetrate
deeper into the atmosphere. Eq.~(\ref{eq:clausius}) implies that
\begin{equation}
{d\ln x\over d\ln P}=\beta\nabla_T -1,
\end{equation}
and hence, on the night side,
\begin{equation}
\ln x(P)=\ln x(\psday) + (\beta\nabla_T-1) \ln(P/\psday),
\end{equation}
assuming that $\nabla_T$ is constant.
Using eq.~(\ref{eq:xnight}), one obtains the condensation pressure
on the night side:
\begin{equation}
{\psnight\over\psday} = \left({\tday\over
\tnight}\right)^{\beta/(\beta\nabla_T -1)}.
\label{eq:psnight}
\end{equation}
Using
$\beta\sim 10$, $\nabla_T\sim 0.15$ and $\tday/\tnight\sim 1.2$, one finds
$P^\star_{\rm night}\sim 38 P^\star_{\rm day}$, a very significant
variation of the condensation pressure. This implies that air flowing
on constant pressure levels around the planet would lead to a rapid
depletion of any condensing species on the day side, compared to what
would be predicted from chemical equilibrium calculations. This can
potentially also remove important absorbing gases from the day side,
as in the case of TiO, which can be removed by CaTiO$_3$ condensation,
or Na, removed by Na$_2$S condensation (Lodders \cite{Lod99}).
Of course, most of the variation depends on the exponential factor
$\beta/(\beta\nabla_T-1)$, which is infinite in the limit when the
atmospheric temperature profile and the condensation profile are
parallel to each other. 

In the discussion, we implicitly assumed $\beta\nabla_T-1>0$; however,
when the atmosphere is close to an isotherm, this factor can become
negative. In this case the day/night effect is even more severe, as
the condensing species is entirely removed from this quasi-isothermal
region.

\def\bi#1#2#3#4#5#6{{#1}, {#6}, {#3} {#4}, #5}

\end{document}